\documentclass[letterpaper,twocolumn,showpacs,
preprintnumbers,amsmath,amssymb,floatfix,nofootinbib]{revtex4}

\usepackage{graphicx}
\usepackage{bm}
\usepackage{url,multirow,xcolor}

\begin{document}

\title{ Many-body correlations of quasiparticle random-phase approximation in nuclear matrix element of neutrinoless double-beta decay}

\author{J.\ Terasaki}
\affiliation{Division of Physics and Center for Computational Sciences, University of Tsukuba, Tsukuba 305-8577, Japan}

\begin{abstract} 
We show that the correlations of the quasiparticle random-phase approximation (QRPA) significantly reduce the nuclear matrix element (NME) of neutrinoless double-beta decay by a new mechanism in the calculation for $^{150}$Nd $\rightarrow$ $^{150}$Sm. This effect is due mainly to the normalization factors of the QRPA ground states included in the overlap of intermediate states, to which the QRPA states based on the initial and final ground states are applied. These normalization factors arise according to the definition of the QRPA ground state as the vacuum of quasibosons. Our NME is close to those of other groups in spite of this new reduction effect because we do not use the proton-neutron pairing interaction usually used for reproducing the experimental NME of the two-neutrino double-beta ($2\nu\beta\beta$) decay. Our method can repeoduce the experimental $2\nu\beta\beta$ NME for $^{150}$Nd $\rightarrow$ $^{150}$Sm with the quenching axial-vector current coupling without approacing the breaking point of the QRPA. The consistency of QRPA approaches taking different virtual paths under the closure approximation is also discussed, and an extension of the QRPA ground state is proposed.
\end{abstract}

%
\maketitle
\section{\label{sec:introduction}Introduction}
Studies of neutrinos have entered a very exciting era. Neutrino oscillation \cite{Fuk98,Ahm02,Egu03,Ali05} provided proof that the neutrino was massive, and most of the neutrino oscillation parameters have been determined, leaving only some uncertainties \cite{For12}. Many experiments to observe neutrinoless double-beta ($0\nu\beta\beta$) decay are in progress, e.g., \cite{Med13}, and are expected to clarify whether the neutrino is a Majorana particle. If this decay is observed, it also implies that the lepton number is not conserved. Further, if the transition probabilities of $0\nu\beta\beta$ decay are measured, they can be used to determine the effective neutrino mass with the help of theoretical calculation of the corresponding transition matrix elements. The determination of the effective neutrino mass by this method is particularly important because neutrino oscillation experiments do not give the absolute values of the neutrino masses. 

One reason for the importance of the neutrino mass is that the neutrino has been assumed to be massless in the standard theory, e.g., \cite{Mat13}. Furthermore, the neutrino affects the fluctuation of the mass distribution in the universe, e.g., \cite{Les06}. (This relation provides us with another  possible method of determining the neutrino mass.) The neutrino also plays an important role in energy and momentum transport in supernova explosions, e.g., \cite{Mez99}. The determination of the neutrino mass is one of the most important subjects in modern physics because the neutrino mass significantly affects 
particle and nuclear physics and astrophysics.  

The challenge for nuclear theory is to calculate the nuclear component of the $0\nu\beta\beta$ transition matrix element, called the nuclear matrix element (NME). This is because all the nuclei that researchers plan to use in experiments have mass number $A\geq 48$ \cite{Med13}; therefore, approximations are essential for obtaining the relevant nuclear wave functions. Several approximate methods are currently used to calculate the NME; unfortunately, however, we are faced with a discrepancy among those NMEs, which vary by a factor of 2$-$3 for more than a dozen decay instances \cite{Fae12J}. 

In this paper, we examine a new mechanism for carrying nuclear many-body correlations to the NME in the quasiparticle random-phase approximation (QRPA) approach. The new mechanism manifests itself in overlap of the two QRPA states obtained on the basis of the initial and final ground states. This overlap is not equal to 1 in the QRPA approach because the components of the nuclear wave functions that do not contribute to the transition matrix element from the initial or final state to intermediate states are not included in the QRPA wave functions. In contrast to transition matrix elements from a ground state to QRPA states, we need the explicit wave functions of the QRPA ground states for calculating the overlaps included in the NME. The QRPA ground state is defined as the vacuum of QRPA quasibosons in our calculations, and the QRPA states are constructed by making creation operators for the QRPA states to act on the QRPA ground states. This study is the first to calculate the overlaps based on the above definition of the QRPA ground states and apply them  to the NME. It will be shown that the normalization factors of the QRPA ground state wave functions significantly reduce the NME.

This paper is organized as follows. Section \ref{sec:NME} shows the basic equations of the NME under the closure approximation. In Sec.~\ref{sec:overlap}, the overlap equations that are used in the calculations in this paper are presented. Section \ref{sec:0vbb_transition_operator_2particle_transfer_me} shows the equations of the $0\nu\beta\beta$ transition matrix elements and matrix elements of two-particle transfer. 
In Sec.~\ref{sec:preparative_calculations}, the results of the calculations required before the NME calculation are described with the computational parameters. Renormalization of the QRPA correlations is also discussed. 
The NME calculation is shown in Sec.~\ref{sec:NME_output}, and the results are compared with the NMEs of other groups. In Sec.~\ref{sec:consistency_QRPA_approach}, the consistency of the QRPA approach is discussed in relation to multiple virtual paths of $0\nu\beta\beta$ decay under the closure approximation. Simplified calculations are shown in Sec.~\ref{sec:simple_test_cal} for showing the validity of our claims. The conclusion and a plan for future work are presented in Sec.~\ref{sec:conclusion}.

\section{\label{sec:NME}NME and virtual paths of $\bm{0\nu\beta\beta}$ decay under closure approximation}
The original equation of the NME, e.g., \cite{Doi85}, is 
\begin{eqnarray}
M^{(0\nu)} &=& \sum_a\sum_{pn}\sum_{p^\prime n^\prime} 
V^{(0\nu)}_{pp^\prime,nn^\prime}(E_a) \nonumber \\
&&\times\langle F(A,Z+2)|c^\dagger_p c_n | a(A,Z+1)\rangle \nonumber \\
&&\times\langle a(A,Z+1)|c^\dagger_{p^\prime}c_{n^\prime} | I(A,Z)\rangle , \label{eq:NME_original}
\end{eqnarray}
where $|I(A,Z)\rangle$ and $|F(A,Z+2)\rangle$ denote the initial and final states of the $0\nu\beta\beta$ decay, respectively ($Z$ is the proton number), and $|a(A,Z+1)\rangle$ denotes the intermediate nuclear state. Further, $\{p,p^\prime\}$ and $\{n, n^\prime \}$ are the proton and neutron single-particle states, respectively, and their creation and annihilation operators are denoted as $\{c^\dagger_p, c^\dagger_n \}$ and $\{c_p, c_n\}$, respectively.   
In this paper, we use the $0\nu\beta\beta$ transition operator consisting of only the Gamow--Teller and Fermi terms; its matrix element is given by 
\begin{equation}
V^{(0\nu)}_{pp^\prime,nn^\prime}(E_a) = V^{GT(0\nu)}_{pp^\prime,nn^\prime}(E_a) +
V^{F(0\nu)}_{pp^\prime,nn^\prime}(E_a), \label{eq:neutrino_potential_me}
\end{equation}
\begin{eqnarray}
\lefteqn{ V^{GT(0\nu)}_{pp^\prime,nn^\prime}(E_a)} \nonumber \\
 &&=\langle pp^\prime | h_+(r_{12},E_a) \bm{\sigma}(1)\cdot{\bm\sigma}(2)
\tau^+(1) \tau^+(2) |nn^\prime \rangle, \label{eq:GT_me} \\
\lefteqn{ V^{F(0\nu)}_{pp^\prime,nn^\prime}(E_a) } \nonumber \\
&&= -(g_V^2/g_A^2) \langle pp^\prime | h_+(r_{12},E_a) \tau^+(1) \tau^+(2) |nn^\prime \rangle, \label{eq:Fermi_me}
\end{eqnarray}
where $\bm{\sigma}(1)$ and $\tau^+(1)$ represent the Pauli spin and charge-changing (neutron$\rightarrow$proton) operators, respectively, and the number in parentheses distinguishes the two particles in the two-body matrix element. Further, $g_V$ and $g_A$ are the vector and axial vector current coupling constants, respectively; $r_{12}$ is a distance variable between the two particles, and $E_a$ is the energy of the intermediate state $a$. We use the approximate neutrino potential \cite{Gre60}
\begin{eqnarray}
h_+(r_{12},E_a) &\simeq& \frac{R}{r_{12}}\frac{2}{\pi}\bigl\{ \sin(\frac{c}{\hbar}\mu_a m_e r_{12})\textrm{ci}(\frac{c}{\hbar}\mu_a m_e r_{12}) \nonumber \\
&&-\cos(\frac{c}{\hbar}\mu_a m_e r_{12})\textrm{si}(\frac{c}{\hbar}\mu_a m_e r_{12}) \bigr\}, \label{eq:neutrino_potential}
\end{eqnarray}
\vspace*{-10pt}
\begin{eqnarray}
\mu_a m_e c^2 &=& E_a - (M_i c^2 + M_f c^2)/2, \label{eq:mume} 
\end{eqnarray}
neglecting the effective neutrino mass relative to the major momentum transfer of the propagating neutrino. The functions ci and si are the cosine and sine integrals, respectively, which are defined as
\begin{eqnarray}
\textrm{si}(x) &=& -\int_x^\infty \frac{\sin(t)}{t}dt, \nonumber\\
\textrm{ci}(x) &=& -\int_x^\infty \frac{\cos(t)}{t}dt,
\end{eqnarray}

\vspace*{5pt}
\noindent
and $M_i$ and $M_f$ indicate 
the masses of the initial and final nuclei, respectively. $R$ is the mean nuclear radius. 

Under the closure approximation, to replace $E_a$ in $M^{(0\nu)}$ [Eq.~(\ref{eq:NME_original})] with an average energy $\bar{E}_a$, 
the NME can be calculated in multiple ways. Some of the possible equations are
\begin{widetext}
\begin{eqnarray}
M^{(0\nu)} &\simeq& \sum_{pn}\sum_{p^\prime n^\prime} 
V^{(0\nu)}_{pp^\prime,nn^\prime}(\bar{E}_a)
\langle F(A,Z+2)|c^\dagger_p c_n c^\dagger_{p^\prime}c_{n^\prime} | I(A,Z)\rangle  \label{eq:NME_closure_no_intermediate} \label{eq:NME_closure_direct} \\
&\simeq& \sum_a\sum_{pn}\sum_{p^\prime n^\prime} 
V^{(0\nu)}_{pp^\prime,nn^\prime}(\bar{E}_a)
\langle F(A,Z+2)|c^\dagger_p c_n | a(A,Z+1)\rangle
\langle a(A,Z+1)|c^\dagger_{p^\prime}c_{n^\prime} | I(A,Z)\rangle  \label{eq:NME_closure_double_beta} \\
&\simeq& \sum_a\sum_{pn}\sum_{p^\prime n^\prime} 
V^{(0\nu)}_{pp^\prime,nn^\prime}(\bar{E}_a)
\langle F(A,Z+2)|c^\dagger_p c^\dagger_{p^\prime} | a(A-2,Z)\rangle
\langle a(A-2,Z)|c_{n^\prime}c_n | I(A,Z)\rangle 
\label{eq:NME_closure_2n_removal_2p_addition} \\
&\simeq& \sum_a\sum_{pn}\sum_{p^\prime n^\prime} 
V^{(0\nu)}_{pp^\prime,nn^\prime}(\bar{E}_a)
\langle F(A,Z+2)|c_{n^\prime}c_n | a(A+2,Z+2)\rangle
\langle a(A+2,Z+2)|c^\dagger_p c^\dagger_{p^\prime} | I(A,Z)\rangle. 
\label{eq:NME_closure_2p_addition_2n_removal}
\end{eqnarray}
The first one is the typical equation used in practical calculations other than the QRPA approach. The last three equations correspond to different virtual paths in the nuclear chart; see Fig.~1 in Ref.~\cite{Ter13}. The spaces of the intermediate states should be such that they cover the space 
obtained by the first (inverse second) step of $0\nu\beta\beta$ decay from the initial (final) state, e.g., $\{c^\dagger_p c_n |I(A,Z)\rangle \}$. In the QRPA approach, (hereafter, all the nuclear states are those of the QRPA), we can use
\begin{eqnarray}
M^{(0\nu)} 
&\simeq& \sum_{a_I a_F}\sum_{pn}\sum_{p^\prime n^\prime} 
V^{(0\nu)}_{pp^\prime,nn^\prime}(\bar{E}_a)
\langle F(A,Z+2)|c^\dagger_p c_n | a_F(A,Z+1)\rangle
\langle a_F(A,Z+1)| a_I(A,Z+1)\rangle \nonumber \\
&&\times\langle a_I(A,Z+1)|c^\dagger_{p^\prime}c_{n^\prime} | I(A,Z)\rangle  \label{eq:NME_closure_double_beta_QRPA}
\end{eqnarray}
\begin{eqnarray}
\hspace{30pt}&\simeq& \sum_{a_I a_F}\sum_{pn}\sum_{p^\prime n^\prime} 
V^{(0\nu)}_{pp^\prime,nn^\prime}(\bar{E}_a)
\langle F(A,Z+2)|c^\dagger_p c^\dagger_{p^\prime} | a_F(A-2,Z)\rangle
\langle a_F(A-2,Z)| a_I(A-2,Z)\rangle \nonumber \\
&&\times\langle a_I(A-2,Z)|c_{n^\prime}c_n | I(A,Z)\rangle 
\label{eq:NME_closure_2n_removal_2p_addition_QRPA} \\
&\simeq& \sum_{a_I a_F}\sum_{pn}\sum_{p^\prime n^\prime} 
V^{(0\nu)}_{pp^\prime,nn^\prime}(\bar{E}_a)
\langle F(A,Z+2)|c_{n^\prime}c_n | a_F(A+2,Z+2)\rangle
\langle a_F(A+2,Z+2)| a_I(A+2,Z+2)\rangle \nonumber \\
&&\times\langle a_I(A+2,Z+2)|c^\dagger_p c^\dagger_{p^\prime} | I(A,Z)\rangle , 
\label{eq:NME_closure_2p_addition_2n_removal_QRPA}
\end{eqnarray}
\end{widetext}
where $a_I$ and $a_F$ denote the QRPA states obtained on the basis of the initial and final ground states, respectively. The usual equation is Eq.~(\ref{eq:NME_closure_double_beta_QRPA}) with the intermediate states obtained by the proton-neutron QRPA. The equations using two sets of intermediate states seem suitable to the QRPA approach because it is easy to calculate the charge-changing, or two-particle transfer, transition matrix elements. In this paper, we use Eq.~(\ref{eq:NME_closure_2n_removal_2p_addition_QRPA}) with the like-particle QRPA because the like-particle QRPA is known to be a good approximation for well-deformed rare-earth nuclei. This 
is one reason for calculating $^{150}$Nd$\rightarrow$$^{150}$Sm in this paper. The choice of either Eq.~(\ref{eq:NME_closure_2p_addition_2n_removal_QRPA}) or Eq. (\ref{eq:NME_closure_2n_removal_2p_addition_QRPA}) is arbitrary. 
An analogous idea for using the two-particle transfer path is discussed in Ref.~\cite{Bar09}  
and also suggested in Ref.~\cite{vog10}. The decay $^{150}$Nd $\rightarrow$ $^{150}$Sm is known to have a large phase-space factor \cite{Kot12,Sto13}, so the reliable NME is highly valuable from an experimental viewpoint.

\section{\label{sec:overlap}Overlap}
In Ref.~\cite{Ter13}, we investigated how to calculate the overlap of two QRPA states based on different nuclei by calculating up to negligible-order terms in the expansion of the overlap with respect to the backward amplitudes of the QRPA solutions.\footnote{The authors of Ref.~\cite{Mus13} write ``the convergence of which (our expansion) may not be fast'' without showing any calculation.} 
Here we note the equations of the overlap that include only the relevant terms and are used in the calculations in this paper.

Hereafter we omit the mass number and proton number of the nuclear states and introduce creation operators of the QRPA states, $O^{I\dagger}_a$ and $O^{F\dagger}_a$:
\begin{eqnarray}
|a_I \rangle &\equiv& |a_I(A-2,Z)\rangle = O^{I\dagger}_a |I\rangle, \nonumber \\
|a_F \rangle &\equiv& |a_F(A-2,Z)\rangle = O^{F\dagger}_a |F\rangle. \label{eq:QRPA_state}
\end{eqnarray}
$O^{^I\dagger}_a$ and $O^{F\dagger}_a$ are expressed using the forward $X^{Ia}_{\mu\nu}$ ($X^{Fa}_{\mu\nu}$) and backward $Y^{Ia}_{-\mu-\nu}$ ($Y^{Fa}_{-\mu-\nu}$) amplitudes of the QRPA solutions as 
\begin{eqnarray}
O^{I\dagger}_a &=& \sum_{\mu<\nu}\bigl( X^{Ia}_{\mu\nu}a^{I\dagger}_\mu a^{I\dagger}_\nu 
-Y^{Ia}_{-\mu-\nu} a^I_{-\nu} a^I_{-\mu} \bigr), \nonumber \\
O^{F\dagger}_a &=& \sum_{\mu<\nu}\bigl( X^{Fa}_{\mu\nu}a^{F\dagger}_\mu a^{F\dagger}_\nu 
-Y^{Fa}_{-\mu-\nu} a^F_{-\nu} a^F_{-\mu} \bigr), \label{eq:QRPA_creation_operators}
\end{eqnarray}
where $-\mu$ indicates that the sign of the $z$ component of the angular momentum of the quasiparticle state $\mu$ is inverted, and these quasiparticle states are ordered to choose $\mu$ and $\nu$ such that 
$\mu < \nu$. 

The QRPA ground states, i.e., the initial and final states of $0\nu\beta\beta$ decay, are defined, e.g., \cite{Rin80}, as 
\begin{eqnarray}
&&O^I_a |I\rangle = 0,\ \nonumber\\
&&O^F_a | F\rangle = 0, \ \ \textrm{for all $a$}, \label{eq:QRPA_ground_state}
\end{eqnarray}
and these ground states can be expressed, e.g., \cite{Bal69}, as
\begin{eqnarray}
|I\rangle &=& \frac{1}{{\cal N}_I} \prod_{K\pi}\exp\left[ \hat{v}^{(K\pi)}_I\right] |i\rangle, \nonumber\\
|F\rangle &=& \frac{1}{{\cal N}_F} \prod_{K\pi}\exp\left[ \hat{v}^{(K\pi)}_F\right] |f\rangle, 
\label{eq:QRPA_ground_states}
\end{eqnarray}
where ${\cal N}_I$ and ${\cal N}_F$ are the normalization factors, and $|i\rangle$ and $|f\rangle$ denote the Hartree--Fock--Bogoliubov (HFB) ground states. $K$ denotes the $z$ component of the angular momentum, and $\pi$ is parity. Throughout this paper, the nuclei under consideration are assumed to have axial and reflection symmetry. We can write $\hat{v}^{(K\pi)}_I$ and $\hat{v}^{(K\pi)}_F$ as
\begin{eqnarray}
\hat{v}^{(K\pi)}_I &=& \sum_{\mu\nu\mu^\prime\nu^\prime} 
C^{(K\pi)I}_{\mu\nu,-\mu^\prime-\nu^\prime} a^{I\dagger}_\mu a^{I\dagger}_\nu a^{I\dagger}_{-\mu^\prime} a^{I\dagger}_{-\nu^\prime}, \nonumber \\
\hat{v}^{(K\pi)}_F &=& \sum_{\mu\nu\mu^\prime\nu^\prime} 
C^{(K\pi)F}_{\mu\nu,-\mu^\prime-\nu^\prime} a^{F\dagger}_\mu a^{F\dagger}_\nu a^{F\dagger}_{-\mu^\prime} a^{F\dagger}_{-\nu^\prime}, \label{eq:v-operators}
\end{eqnarray}
\begin{equation}
a^I_\mu |i\rangle = a^F_\mu |f\rangle = 0. \label{eq:qp_operators_def}
\end{equation}
By using the quasiboson approximation, e.g., \cite{Row68}, to ignore the exchange terms,  
$C^{(K\pi)I}_{\mu\nu,-\mu^\prime-\nu^\prime}$ and $C^{(K\pi)F}_{\mu\nu,-\mu^\prime-\nu^\prime}$, which are called the correlation coefficients \cite{Ull72}, are obtained: 
\begin{eqnarray}
C^{(K\pi)I}_{\mu\nu,-\mu^\prime -\nu^\prime} &=& \frac{1}{1+\delta_{K0}} \sum_a 
Y^{Ia\ast}_{-\mu^\prime -\nu^\prime} \left( \frac{1}{X^{(K\pi)I\ast}} \right)_{a,\mu\nu}, \nonumber \\
C^{(K\pi)F}_{\mu\nu,-\mu^\prime -\nu^\prime} &=& \frac{1}{1+\delta_{K0}} \sum_a 
Y^{Fa\ast}_{-\mu^\prime -\nu^\prime} \left( \frac{1}{X^{(K\pi)F\ast}} \right)_{a,\mu\nu},
\nonumber \\
\label{eq:correlation_coefficient}
\end{eqnarray}
where the matrix $X^{(K\pi)I\ast}$ ($X^{(K\pi)F\ast}$), which consists of $X^{Ia\ast}_{\mu\nu}$ ($X^{Fa\ast}_{\mu\nu}$), is used, and the summations include only those QRPA solutions $a$ having $(K\pi)$. See Sec.~II~C of Ref.~\cite{Ter13} for the definition of the matrix notation.

The correlation coefficients are proportional to the backward amplitudes of the QRPA solutions and bring the QRPA many-body correlations to the QRPA ground states. Note that $\hat{v}^{(K\pi)}_I$ and $\hat{v}^{(K\pi)}_F$ 
are operators of the $z$ component of the angular momentum with 0 and positive parity because we always consider only even-even nuclei; $(K\pi)$ implies that these operators consist of the products of two operators with $(K\pi)$ and $(-K\pi)$. 

The overlap is expanded with respect to $\hat{v}^{(K\pi)}_I$ and $\hat{v}^{(K\pi)}_F$: 
\begin{eqnarray}
\langle a_F | a^\prime_I \rangle 
&\simeq& \frac{1}{{\cal N}_I {\cal N}_F} \biggl\{ \langle f | O^F_a O^{I\dagger}_{a^\prime} | i \rangle 
\nonumber\\
&&+\langle f | \hat{v}^{(K\pi)\dagger}_F O^F_a O^{I\dagger}_{a^\prime} |i\rangle \nonumber \\
 &&+\langle f | O^F_a O^{I\dagger}_{a^\prime} \hat{v}^{(K\pi)}_I |i\rangle \biggr\},  \label{eq:overlap_expansion}
\end{eqnarray}
 where $(K\pi)$ is that of the QRPA states $a_F$ and $a^\prime_I$, and 
\begin{eqnarray}
{\cal N}_I &=& \langle i | \prod_{K\pi} \exp{\bigl[\hat{v}^{(K\pi)\dagger}_I\bigr]} \exp{\bigl[\hat{v}^{(K\pi)}_I\bigr]} |i\rangle^{1/2} \nonumber\\
&\simeq& \biggl[ 1 + \sum_{K\pi}\biggl\{ \sum_{n=1}^5\frac{1}{n!}(1+\delta_{K0})^n \bigl\{\textrm{Tr}(C^{(K\pi)I} C^{(K\pi)I\dagger})\bigr\}^n \nonumber\\
&&+\frac{1}{2}(1+7\delta_{K0})\textrm{Tr}(C^{(K\pi)I}C^{(K\pi)I\dagger})^2 \biggl\} \biggl]^{1/2}, \nonumber 
\end{eqnarray}
\vspace{10pt}
\begin{eqnarray}
{\cal N}_F &=& \langle f | \prod_{K\pi} \exp{\bigl[\hat{v}^{(K\pi)\dagger}_F\bigr]} \exp{\bigl[\hat{v}^{(K\pi)}_F\bigr]} |f\rangle^{1/2} \nonumber\\
&\simeq& \biggl[ 1 + \sum_{K\pi}\biggl\{ \sum_{n=1}^5\frac{1}{n!}(1+\delta_{K0})^n 
\nonumber\\
&&\times\bigl\{\textrm{Tr}(C^{(K\pi)F} C^{(K\pi)F\dagger})\bigr\}^n \nonumber\\
&&\hspace{-8pt}+\frac{1}{2}(1+7\delta_{K0}) \textrm{Tr}(C^{(K\pi)F}C^{(K\pi)F\dagger})^2 \biggl\} \biggl]^{1/2}. \label{eq:normalization_factor}
\end{eqnarray}
$C^{(K\pi)I}$ and $C^{(K\pi)F}$ are matrices consisting of $C^{(K\pi)I}_{\mu\nu,-\mu^\prime-\nu^\prime}$ and $C^{(K\pi)I}_{\mu\nu,-\mu^\prime-\nu^\prime}$, respectively. 
If the good quantum numbers of the two QRPA states differ, the overlap vanishes. The terms proportional to 
$\{\textrm{Tr}(C^{(K\pi)I}C^{(K\pi)I\dagger})\}^n$ are calculated up to $n=5$ in Eq.~(\ref{eq:normalization_factor}) 
because it is easy to calculate them (the analytical expression including all orders can also be used); however, the terms with $n>2$ are very small in our numerical calculation. 

By using Eq.~(\ref{eq:QRPA_creation_operators}) and the correlation coefficients, the components of  Eq.~(\ref{eq:overlap_expansion}) are obtained:
\begin{eqnarray}
\lefteqn{ \langle f| O^{F}_a O^{I\dagger}_{a^\prime} |i\rangle } \nonumber \\
&&= \sum_{\mu<\nu} X^{Fa\ast}_{\mu\nu} \sum_{\mu^\prime \nu^\prime} X^{Ia^\prime}_{\mu^\prime \nu^\prime}
\langle f| a^F_\nu a^F_\mu a^{I\dagger}_{\mu^\prime} a^{I\dagger}_{\nu^\prime} |i\rangle, 
\label{eq:0th_unnormalized_overlap}
\end{eqnarray}
\begin{widetext}
\begin{eqnarray}
\lefteqn{\langle f| \hat{v}^{(K\pi)\dagger}_F O^F_a O^{I\dagger}_{a^\prime} |i\rangle } \nonumber \\
&=& \sum_{\mu\nu\mu^\prime\nu^\prime}\sum_{\mu_1<\nu_1}\sum_{\mu_2<\nu_2}
C^{(K\pi)F\ast}_{\mu\nu,\mu^\prime\nu^\prime} X^{Fa\ast}_{\mu_1\nu_1} X^{Ia^\prime}_{\mu_2\nu_2}
\langle f| a^F_{\nu^\prime} a^F_{\mu^\prime} a^F_\nu a^F_\mu a^F_{\nu_1} a^F_{\mu_1} a^{I\dagger}_{\mu_2} a^{I\dagger}_{\nu_2} |i\rangle \nonumber \\
&&-\sum_{\mu\nu}\sum_{\mu_1<\nu_1}\sum_{\mu_2<\nu_2}
Y^{Fa\ast}_{-\mu_1-\nu_1} X^{Ia^\prime}_{\mu_2\nu_2} \bigl\{
C^{(K\pi)F\ast}_{-\nu_1-\mu_1,\mu\nu} -C^{(K\pi)F\ast}_{-\mu_1-\nu_1,\mu\nu}
+C^{(K\pi)F\ast}_{\mu\nu,-\nu_1-\mu_1} -C^{(K\pi)F\ast}_{\mu\nu,-\mu_1-\nu_1}
+C^{(K\pi)F\ast}_{-\nu_1\nu,-\mu_1\mu} \nonumber \\
&&-C^{(K\pi)F\ast}_{-\mu_1\nu,-\nu_1\mu}
-C^{(K\pi)F\ast}_{-\nu_1\nu,\mu-\mu_1} +C^{(K\pi)F\ast}_{-\mu_1\nu,\mu-\nu_1} 
+C^{(K\pi)F\ast}_{\mu-\nu_1,-\mu_1\nu} -C^{(K\pi)F\ast}_{\mu-\mu_1,-\nu_1\nu}
-C^{(K\pi)F\ast}_{\mu-\nu_1,\nu-\mu_1} +C^{(K\pi)F\ast}_{\mu-\mu_1,\nu-\nu_1} \bigl\} \nonumber \\
&&\times \langle f| a^F_\mu a^F_\nu a^{I\dagger}_{\mu_2} a^{I\dagger}_{\nu_2} | i\rangle, 
\label{eq:1st-1_unnormalized_overlap}
\end{eqnarray}
\begin{eqnarray}
\lefteqn{ \langle f| O^F_a O^{I\dagger}_{a^\prime} \hat{v}^{(K\pi)}_I | i\rangle } \nonumber \\
&=& \sum_{\mu<\nu} \sum_{\mu^\prime<\nu^\prime}\sum_{\mu_1\nu_1}\sum_{\mu_2\nu_2}
X^{Fa\ast}_{\mu\nu} X^{Ia^\prime}_{\mu^\prime\nu^\prime} 
C^{(K\pi)I}_{\mu_1\nu_1,\mu_2\nu_2} 
\langle f| a^F_\nu a^F_\mu a^{I\dagger}_{\mu^\prime}a^{I\dagger}_{\nu^\prime}
a^{I\dagger}_{\mu_1} a^{I\dagger}_{\nu_1} a^{I\dagger}_{\mu_2} a^{I\dagger}_{\nu_2} |i\rangle \nonumber \\
&& -\sum_{\mu<\nu} \sum_{\mu^\prime<\nu^\prime} \sum_{\mu_1\mu_2}
X^{Fa\ast}_{\mu\nu} Y^{Ia^\prime}_{-\mu^\prime-\nu^\prime} \bigl\{
-C^{(K\pi)I}_{\mu_1\mu_2,-\nu^\prime-\mu^\prime} +C^{(K\pi)I}_{\mu_1\mu_2,-\mu^\prime-\nu^\prime}
-C^{(K\pi)I}_{-\nu^\prime-\mu\prime,\mu_1\mu_2} +C^{(K\pi)I}_{-\mu^\prime-\nu^\prime,\mu_1\mu_2}
-C^{(K\pi)I}_{\mu_1-\mu^\prime,\mu_2-\nu^\prime} \nonumber \\
&&+C^{(K\pi)I}_{\mu_1-\nu^\prime,\mu_2-\mu^\prime} +C^{(K\pi)I}_{\mu_1-\mu^\prime,-\nu^\prime\mu_2}
-C^{(K\pi)I}_{\mu_1-\nu^\prime,-\mu^\prime\mu_2} +C^{(K\pi)I}_{-\mu^\prime\mu_1,\mu_2-\nu^\prime} 
-C^{(K\pi)I}_{-\nu^\prime\mu_1,\mu_2-\mu^\prime} -C^{(K\pi)I}_{-\mu^\prime\mu_1,-\nu^\prime\mu_2} \nonumber \\
&&+C^{(K\pi)I}_{-\nu^\prime\mu_1,-\mu^\prime\mu_2} \bigr\}
\langle f| a^F_\nu a^F_\mu a^{I\dagger}_{\mu_1} a^{I\dagger}_{\mu_2} | i\rangle .
\label{eq:1st-2_unnormalized_overlap}
\end{eqnarray}
\end{widetext}
Equations (\ref{eq:overlap_expansion})$-$(\ref{eq:1st-2_unnormalized_overlap}) are used to compute the overlap. 
The explicit equation of  
$\langle f|a^F_\nu a^F_\mu a^{I\dagger}_{\mu^\prime} a^{I\dagger}_{\nu^\prime} |i\rangle$  
is given in Ref.~\cite{Ter13}.\footnote{The superscripts $I$ in the second line of Eq.~(49) in Ref.~\cite{Ter13} should read $F$.} 

The justification of the approximations used in the above equations  
is discussed in detail in Ref.~\cite{Ter13}. Here we make a few remarks on those approximations. The unnormalized overlap [the right-hand side of Eq.~(\ref{eq:overlap_expansion}) except for the normalization factors] is truncated at the first order with respect to $C^{(K\pi)I}$ or $C^{(K\pi)F}$, whereas ${\cal N}_i^2$ and ${\cal N}_F^2$ are expanded up to the fourth order (and partially more). This difference arises from the special characteristic that the initial and final states are the ground states of different nuclei; the high-energy excitation components of the operators in Eq.~(\ref{eq:overlap_expansion}) that do not affect the Fermi surface region make almost no contribution to the unnormalized overlap. The normalization factors, however, do not have this characteristic. 

If the values of $(K\pi)$ in $\hat{v}^{(K\pi)\dagger}_F$ and $\hat{v}^{(K\pi)}_I$ in Eq.~(\ref{eq:overlap_expansion}) are not equal to those of $O^{F\dagger}_a$ and $O^{I\dagger}_{a^\prime}$, the contributions of those terms are very small. 
This has been checked in the test calculation in Ref.~\cite{Ter13}; it is also an expected property because operators with different good quantum numbers commute with each other in the QRPA order, and roughly the only difference between $|f\rangle $ and $|i\rangle$ is the configuration at the Fermi surface.

In Eq.~(\ref{eq:normalization_factor}), the exchange terms that are not expressed in the form of the trace of matrix multiplication are neglected. The exchange terms are smaller than the terms included in Eq.~(\ref{eq:normalization_factor}) called the quasiboson terms; this was checked in the test calculation in Ref.~\cite{Ter13}.  
Note that the exchange terms have more selection rules based on the good quantum numbers than the quasiboson terms do [see Eqs.~(\ref{eq:1st-1_unnormalized_overlap}) and (\ref{eq:1st-2_unnormalized_overlap})]; therefore, the number of exchange terms is much smaller than the number of quasiboson terms. This is the main property enabling the approximation to neglect the exchange terms (there is no reason that each exchange term is significantly larger than the quasiboson terms). 

The generalized expectation values of the product of the many quasiparticle creation and annihilation 
operators in Eqs.~(\ref{eq:1st-1_unnormalized_overlap}) and (\ref{eq:1st-2_unnormalized_overlap}) 
are calculated systematically using the algorithm of the proof of the generalized Wick's theorem \cite{Bal69}   
without distinguishing the direct and exchange terms; thus, these equations include both types of terms.

\section{\label{sec:0vbb_transition_operator_2particle_transfer_me}matrix elements of $\bm{0\nu\beta\beta}$ transition operator and two-particle transfer}
\subsection{\label{subsec:0vbb_transition_operator_me} Matrix elements of $\bm{0\nu\beta\beta}$ transition operator }
We showed the equation of the $0\nu\beta\beta$ transition operator used in this paper in Sec.~\ref{sec:NME}. In this section, we show the equation of the matrix element of that operator for computation. The quasiparticle and single-particle wave functions are always numerically expressed in a B-spline mesh, e.g., \cite{Boo78,Nur89}, in a cylindrical box in our calculations; therefore, the wave functions of the relative and center-of-mass motions are not trivial. Thus, we calculate the two-body matrix elements of the $0\nu\beta\beta$ transition operator using the product wave functions of the two single-particle states in the laboratory frame. Note that $p$ and $p^\prime$ ($n$ and $n^\prime$) in the equations for $M^{(0\nu)}$ in Sec.~\ref{sec:NME} cover all the proton (neutron) states, so both the direct and exchange matrix elements are included; see Eq.~(\ref{eq:NME_closure_direct}). 

The single-particle wave functions used in our calculations are expressed as 
\begin{eqnarray}
\mathit{\Psi}_i(\bm{r}_1) = \frac{1}{ \sqrt{2\pi} }\sum_{\sigma=\pm 1/2} \mathcal{F}_i(\sigma;z,\rho)
e^{i(j^z_i-\sigma)\phi} |\sigma\rangle, \label{eq:single-particle_wf}
\end{eqnarray}
 taking into account the axial symmetry of the nuclei considered. The label $i$ implies $(\pi_i,j^z_i,n_i)$, where $j^z_i$ is the $z$ component of the angular momentum of the single-particle state, and 
$n_i$ is a label distinguishing single-particle states in the $(\pi_i,j^z_i)$ subspace.
$\sigma$ is the $z$ component of the spin, and $|\sigma\rangle$ is a spin wave function. 
The variables $(z,\rho,\phi)$ are the cylindrical coordinates 
($\rho$ and $\phi$ represent the radius and angle in the $xy$ plane, respectively).
The function $\mathcal{F}_i(\sigma;z,\rho)$ is treated numerically in our calculations and has reflection symmetry:
\begin{equation}
\mathcal{F}_i(\sigma;-z,\rho) = (-)^{l^z_i} \pi_i \mathcal{F}_i(\sigma;z,\rho),\ \ l^z_i = j^z_i -\sigma. 
\label{eq:parity_symmetry_F}
\end{equation}
$\mathcal{F}_i(\sigma;z,\rho)$ is real in computation without losing generality. 
By using the wave function of Eq.~(\ref{eq:single-particle_wf}), $V^{GT(0\nu)}_{pp^\prime, nn^\prime}(\bar{E}_a)$ [Eq.~(\ref{eq:GT_me}) with the closure approximation] can be written as 
\begin{eqnarray}
\lefteqn{ V^{GT(0\nu)}_{pp^\prime, nn^\prime}(\bar{E}_a) } \nonumber \\
&=& \frac{1}{(2\pi)^2}\int^\infty_0 d\rho_1\rho_1 \int^\infty_0 d\rho_2 \rho_2
\int^\infty_0 dz_1 \int^\infty_0 dz_2 \nonumber \\
&&\biggl[ \sum_{\sigma_p,\sigma_{p^\prime}=\pm1/2} \mathcal{F}^\ast_p(\sigma_p;z_1,\rho_1)
\mathcal{F}^\ast_{p^\prime}(\sigma_{p^\prime};z_2,\rho_2)
\mathcal{F}_n(\sigma_p;z_1,\rho_1) \nonumber \\
&&\times\mathcal{F}_{n^\prime}(\sigma_{p^\prime};z_2,\rho_2)
\Bigl\{ \mathcal{I}(z_1,\rho_1,z_2,\rho_2;-j^z_p+j^z_n) \nonumber \\
&&-(-)^{j^z_{p^\prime}+j^z_{n^\prime}}\pi_{p^\prime}\pi_{n^\prime}
\mathcal{I}(z_1,\rho_1,-z_2,\rho_2;-j^z_p+j^z_n) \Bigr\}(-)^{\sigma_p-\sigma_{p^\prime}} \nonumber \\
&&+2\sum_{\sigma_p}\mathcal{F}_p^\ast(\sigma_p;z_1,\rho_1)
\mathcal{F}_{p^\prime}^\ast(-\sigma_p;z_2,\rho_2) \mathcal{F}_n(-\sigma_p;z_1,\rho_1)\nonumber \\
&&\times\mathcal{F}_{n^\prime}(\sigma_p;z_2,\rho_2)
\Bigl\{ \mathcal{I}(z_1,\rho_1,z_2,\rho_2;-j^z_p+j^z_n+2\sigma_p) \nonumber \\
&&+(-)^{j^z_{p^\prime}+j^z_{n^\prime}}\pi_{p^\prime}\pi_{n^\prime}
\mathcal{I}(z_1,\rho_1,-z_2,\rho_2;-j^z_p+j^z_n+2\sigma_p) \Bigr\} \biggr] \nonumber \\
&&\times 2\delta_{\pi_p\pi_{p^\prime}\pi_n\pi_{n^\prime},1}
\delta_{j^z_p+j^z_{p^\prime},j^z_n+j^z_{n^\prime}}, \label{eq:GT_me_1}
\end{eqnarray} 
\begin{eqnarray}
\mathcal{I}(z_1,\rho_1,z_2,\rho_2;I) = 4\pi\int_0^\pi d\mathit{\Phi} \cos (I\mathit{\Phi})
h_+(r_{12},\bar{E}_a), \label{eq:mathkarI}
\end{eqnarray}
where $I$ is an integer. Note that $r_{12}$ depends on $\mathit{\Phi}$, as shown by
\begin{eqnarray}
r_{12} = \bigl\{ \rho_1^2+\rho_2^2-2\rho_1\rho_2 \cos \mathit{\Phi}+(z_1-z_2)^2 \bigr\}^{1/2}.
\label{eq:r12}
\end{eqnarray}
Equation (\ref{eq:neutrino_potential}) with $R=1.1A^{1/3}$ fm is used for the neutrino potential. 
The matrix element of the Fermi operator is given by 
\begin{eqnarray}
\lefteqn{ V^{F(0\nu)}_{pp^\prime, nn^\prime}(\bar{E}_a) } \nonumber \\
&=& -\frac{g_V^2}{g_A^2}\frac{1}{(2\pi)^2} \int_0^\infty d\rho_1\rho_1
\int_0^\infty d\rho_2\rho_2 \int_0^\infty dz_1\int_0^\infty dz_2  \nonumber \\
&&\sum_{\sigma_p\sigma_{p^\prime}}
\mathcal{F}_{p}^\ast(\sigma_p;z_1,\rho_1) \mathcal{F}_{p^\prime}^\ast(\sigma_{p^\prime};z_2,\rho_2) \mathcal{F}_n(\sigma_p;z_1,\rho_1) \nonumber \\
&&\times \mathcal{F}_{n^\prime}(\sigma_{p^\prime};z_2,\rho_2)
\Bigl\{ \mathcal{I}(z_1,\rho_1,z_2,\rho_2;-j^z_p+j^z_n) \nonumber \\
&&+(-)^{j^z_{p^\prime}+j^z_{n^\prime}} (-)\pi_{p^\prime}\pi_{n^\prime} 
\mathcal{I}(z_1,\rho_1,-z_2,\rho_2;-j^z_p+j^z_n) \Bigr\} \nonumber \\
&&\times 2\delta_{\pi_p\pi_{p^\prime}\pi_n\pi_{n^\prime},1}
\delta_{j^z_{p^\prime}+j^z_p,j^z_n+j^z_{n^\prime}}.
\label{eq:Fermi_me_1}
\end{eqnarray}

\begin{figure}[t]
\vspace{11pt}
\includegraphics[width=6cm]{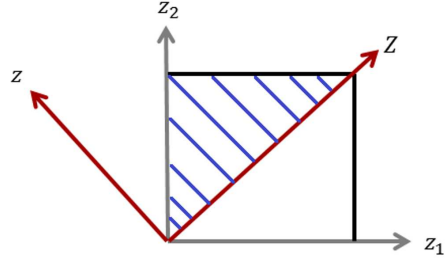}
\caption{ \label{fig:integral_region} (Color online) Integral region (inside the square) and integral paths (the lines from $z$ = 0 to the left or top boundary of the square) for calculating the matrix elements of the $0\nu\beta\beta$ transition operator. Integral with respect to $Z$ is made after the integral with respect to $z$. The same geometry is applied to a $(\rho_1,\rho_2)$ plane.}
\end{figure}

Obtaining the equation for computation requires one more step because the neutrino potential has 
a singularity. We introduce the coordinates $(P,\rho)$ and $(Z,z)$, which are obtained from $(\rho_1,\rho_2)$ and $(z_1,z_2)$, respectively, by a rotation by $\pi/4$ (see Fig.~\ref{fig:integral_region}):
\begin{eqnarray}
&& 
P = \displaystyle{ \frac{1}{\sqrt{2}} }(\rho_1 + \rho_2), \ \ 
\rho = \displaystyle{ \frac{1}{\sqrt{2}} } (-\rho_1 + \rho_2) , \nonumber\\
&&
Z = \frac{1}{\sqrt{2}} (z_1 + z_2), \ \ 
z = \frac{1}{\sqrt{2}} (-z_1 + z_2). \label{eq:PrhoZz}
\end{eqnarray}
The singularity of the neutrino potential occurs along $z_1 = z_2$ and $\rho_1 = \rho_2$ when $\mathit{\Phi}=0$ [see Eqs.~(\ref{eq:neutrino_potential}) and (\ref{eq:r12})]. We can integrate the integrand of Eqs.~(\ref{eq:GT_me_1}) and (\ref{eq:Fermi_me_1}) from $z=0$ to the left or top boundary of the square in Fig.~\ref{fig:integral_region} using the Gauss--Legendre quadrature. 
The $z$ coordinates of this boundary are denoted by the function $B_1(Z)$ in the equation below.
The integral in the region of $z<0$ can be handled analogously. The same geometry can be applied to the integral in the $(\rho_1,\rho_2)$ plane, and the upper boundary of the $\rho$ integral is denoted by $B_2(P)$. In our computation, the data for the single-particle wave functions are provided on a B-spline mesh, as mentioned above; therefore, we obtain the values of the wave functions on the Gauss--Legendre mesh points using a B-spline interpolation formula \cite{Boo78,Nur89}. 

On the basis of the above method, we can use the following equation for the Gamow--Teller matrix element:
\begin{widetext}
\begin{eqnarray}
\lefteqn{ V^{GT(0\nu)}_{pp^\prime, nn^\prime}(\bar{E}_a) } \nonumber \\
&=& \frac{1}{(2\pi)^2} \int_0^\infty dZ \int_0^\infty dP \int_0^{B_1(Z)} dz \int_0^{B_2(P)} d\rho
\rho_1\rho_2\biggl[ \sum_{\sigma_p,\sigma_{p^\prime}=\pm 1/2}\Bigl\{
\mathcal{F}^\ast_p(\sigma_p;z_1,\rho_1)
\mathcal{F}^\ast_{p^\prime}(\sigma_{p^\prime};z_2,\rho_2) 
\mathcal{F}_n(\sigma_p;z_1,\rho_1) \mathcal{F}_{n^\prime}(\sigma_{p^\prime};z_2,\rho_2) \nonumber \\
&&+\mathcal{F}^\ast_p(\sigma_p;z_2,\rho_1) \mathcal{F}^\ast_{p^\prime}(\sigma_{p^\prime};z_1,\rho_2) \mathcal{F}_n(\sigma_p;z_2,\rho_1) 
\mathcal{F}_{n^\prime}(\sigma_{p^\prime};z_1,\rho_2)
+\mathcal{F}^\ast_p(\sigma_p;z_1,\rho_2) 
\mathcal{F}^\ast_{p^\prime}(\sigma_{p^\prime};z_2,\rho_1) 
\mathcal{F}_n(\sigma_p;z_1,\rho_2) \nonumber \\
&&\times\mathcal{F}_{n^\prime}(\sigma_{p^\prime};z_2,\rho_1)
+\mathcal{F}^\ast_p(\sigma_p;z_2,\rho_2) \mathcal{F}^\ast_{p^\prime}(\sigma_{p^\prime};z_1,\rho_1) \mathcal{F}_n(\sigma_p;z_2,\rho_2)
\mathcal{F}_{n^\prime}(\sigma_{p^\prime};z_1,\rho_1) 
\Bigr\}
\Bigl\{\mathcal{I}(\sqrt{2}z,\rho_1,0,\rho_2;-j^z_p+j^z_n) \nonumber \\
&&-(-)^{j^z_{p^\prime}+j^z_{n^\prime}}\pi_{p^\prime}\pi_{n^\prime}
\mathcal{I}(\sqrt{2}Z,\rho_1,0,\rho_2;-j^z_p
+j^z_n) \Bigr\}(-)^{\sigma_p-\sigma_{p^\prime}} \nonumber \\
&&+2\sum_{\sigma_p}\Bigl\{
\mathcal{F}^\ast_p(\sigma_p;z_1,\rho_1) \mathcal{F}^\ast_{p^\prime}(-\sigma_p;z_2,\rho_2) 
\mathcal{F}_n(-\sigma_p;z_1,\rho_1) 
\mathcal{F}_{n^\prime}(\sigma_p;z_2,\rho_2)
+\mathcal{F}^\ast_p(\sigma_p;z_2,\rho_1) 
\mathcal{F}^\ast_{p^\prime}(-\sigma_p;z_1,\rho_2) \nonumber \\
&&\times \mathcal{F}_n(-\sigma_p;z_2,\rho_1) \mathcal{F}_{n^\prime}(\sigma_p;z_1,\rho_2)
+\mathcal{F}^\ast_p(\sigma_p;z_1,\rho_2) \mathcal{F}^\ast_{p^\prime}(-\sigma_p;z_2,\rho_1) \mathcal{F}_n(-\sigma_p;z_1,\rho_2) 
\mathcal{F}_{n^\prime}(\sigma_p;z_2,\rho_1) \nonumber \\
&&+\mathcal{F}^\ast_p(\sigma_p;z_2,\rho_2) 
\mathcal{F}^\ast_{p^\prime}(-\sigma_p;z_1,\rho_1) 
\mathcal{F}_n(-\sigma_p;z_2,\rho_2)
\mathcal{F}_{n^\prime}(\sigma_p;z_1,\rho_1) \Bigr\} 
\Bigl\{ \mathcal{I}(\sqrt{2}z,\rho_1,0,\rho_2;-j^z_p+j^z_n+2\sigma_p)
\nonumber \\
&&+(-)^{j^z_{p^\prime}+j^z_{n^\prime}}\pi_{p^\prime}\pi_{n^\prime}
\mathcal{I}(\sqrt{2}Z,\rho_1,0,\rho_2;-j^z_p+j^z_n+2\sigma_p) \Bigr\} \biggr]
2 \delta_{\pi_p\pi_{p^\prime},\pi_n\pi_{n^\prime}}
\delta_{j^z_p+j^z_{p^\prime},j^z_n+j^z_{n\prime}}, \label{eq:GT_me_2}
\end{eqnarray}
\end{widetext}
where $\rho_1$ and $\rho_2$ ($z_1$ and $z_2$) are functions of $P$ and $\rho$ ($Z$ and $z$) [see Eq.~(\ref{eq:PrhoZz})]. The corresponding equation for the matrix element of the Fermi operator can be derived analogously.

\begin{figure}
\includegraphics[width=6cm]{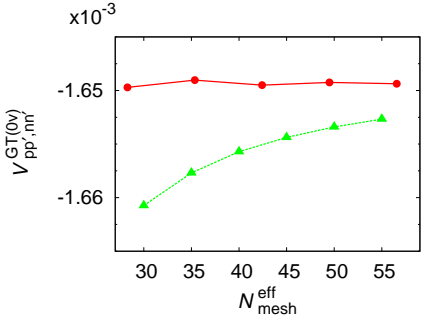}
\caption{ \label{fig:arrange_hplsgt0_comp_0} (Color online) $V^{GT(0\nu)}_{pp^\prime,nn^\prime}(\bar{E}_a)$, an arbitrarily chosen non-negligible one, as a function of effective number of mesh points per dimension $N^\textrm{eff}_\textrm{mesh}$. [$(N^\textrm{eff}_\textrm{mesh})^2$ is equal to the number of mesh points for the $(z_1, z_2)$ plane, and this number is equal to that for the $(\rho_1,\rho_2)$ plane.] 
This test calculation was performed using the setup for $^{26}$Mg$\rightarrow$$^{26}$Si used in Ref.~\cite{Ter13} [max$(z_1)$ = max$(\rho_1)$ = 10 fm]. The circles  show the result using the scheme shown in Fig.~\ref{fig:integral_region}, and the triangles  show the result using the integral paths from an edge of the square in Fig.~\ref{fig:integral_region} to the edge at the opposite side, including the singular points. In the calculations for $^{150}$Nd$\rightarrow$$^{150}$Sm, it is difficult to use $N^\textrm{eff}_\textrm{mesh}$ $\agt$ 50 because too much memory is required.}
\end{figure}
The integral intervals are separated in such a way that the singularity of the integrand is set at the edge, and the values of the integrand at the edge are not used in the Gaussian quadrature.  
The advantage of this method is clear when it is compared to calculations using integral paths not separated at the singular points, as shown in Fig.~\ref{fig:arrange_hplsgt0_comp_0}.
A multipole-multipole expansion \cite{Mus13} can also be used to avoid the difficulty arising from the singularity.

\subsection{\label{subsec:2particle_transfer_me} Matrix elements of two-particle transfer}
As mentioned in Sec.~\ref{sec:NME}, we use Eq.~(\ref{eq:NME_closure_2n_removal_2p_addition_QRPA}) in this paper.
The necessary transition matrix elements in this equation are calculated according to 
\begin{eqnarray}
\langle a_I|c_{-n^\prime} c_{-n} |I\rangle &=&
( -X^{Ia\ast}_{nn^\prime} + X^{Ia\ast}_{n^\prime n} )
s_{n^\prime} v_{n^\prime} s_n v_n  \nonumber \\
&& + (Y^{Ia\ast}_{-n-n^\prime} - Y^{Ia\ast}_{-n^\prime -n}) u_{n^\prime} u_n , 
\label{eq:2n_removal_tr_me} 
\end{eqnarray}
\begin{equation}
s_n = j^z_{n}/\left| j^z_{n} \right| . \label{eq:phase_convention}
\end{equation}
We use the canonical basis \cite{Rin80} associated with the initial HFB ground state for the neutron single-particle states in this equation. The factors $u_n$ and $v_n$ are transition matrix elements defined in the transformation from the canonical to canonical quasiparticle basis.  
Equation (\ref{eq:phase_convention}) is our phase convention. For the proton single-particle states
in $\langle F| c^\dagger_p c^\dagger_{p^\prime} | a_F \rangle$, we use the canonical basis associated with the final HFB ground state. Thus, the wave functions of the protons and neutrons in 
the equations for $V^{GT(0\nu)}_{pp^\prime,nn^\prime}$ and $V^{F(0\nu)}_{pp^\prime,nn^\prime}$ are 
associated with different HFB states. These bases should be sufficiently large.

The final stage of the calculation of the NME is to obtain the trace of the product of the four 
matrices; see Eq.~(\ref{eq:NME_closure_2n_removal_2p_addition_QRPA}).

\section{\label{sec:preparative_calculations} Preparatory calculations}
\subsection{\label{subsec:result_HFB} HFB ground states}
The Skyrme (SkM$^\ast$ \cite{Bar82}) and volume pairing \cite{Dob96} energy density functionals are used in our calculations. The strengths of the volume pairing energy density functional are $-218.521$ MeV fm$^3$ (protons) and $-176.364$ MeV fm$^3$ (neutrons) for $^{150}$Nd and $-218.521$ MeV fm$^3$ (protons) and $-181.655$ MeV fm$^3$ (neutrons) for $^{150}$Sm. These strengths are adjusted so as to reproduce the experimental data obtained from the masses \cite{Aud03} using the three-point formula \cite{Boh69} with deviations of less than 200 keV in the HFB calculations; see Table~\ref{tab:pairing_gaps}. We use the HFB code explained in Refs.~\cite{Ter03,Obe03,Bla05,Obe07}.
The radius and height ($z>0$) of the cylindrical box used in our HFB calculations are both 20 fm, and 42 B-spline mesh points are used per dimension. 
The cutoff quasiparticle energy is 60 MeV. 
We obtained quadrupole deformations $\beta$ of 0.279 for $^{150}$Nd and 0.209 for $^{150}$Sm. 
Two sets of experimental data for the deformations are known: ($\beta$ for $^{150}$Nd, $\beta$ for $^{150}$Sm) = [0.367(86), 0.230(30)] \cite{Lal99} and [0.2853(21), 0.1931(21)] \cite{Ram01}. 
Our values are closer to the latter. 
The total root-mean-square radius of the HFB solutions is $\simeq$5.0 fm for both $^{150}$Nd and $^{150}$Sm. 
\begin{table}
\caption{\label{tab:pairing_gaps} Average pairing gaps of protons ($\mathit{\Delta}_p$) and neutrons
 ($\mathit{\Delta}_n$) in the HFB calculations and those obtained from the mass data using the three-point formula ($\mathit{\Delta}^\textrm{exp}_p$ for protons and $\mathit{\Delta}^\textrm{exp}_n$ for neutrons).}
\begin{ruledtabular}
\begin{tabular}{ccccc}
Nucleus & $\mathit{\Delta}_p$ (MeV) & $\mathit{\Delta}_n$ (MeV) & 
$\mathit{\Delta}^\textrm{exp}_p$ (MeV) & $\mathit{\Delta}^\textrm{exp}_n$ (MeV) \\
\colrule
$^{150}$Nd & 1.494 & 0.925 & 1.464 & 1.023 \\
$^{150}$Sm & 1.869 & 1.058 & 1.692 & 1.195
\end{tabular}
\end{ruledtabular}
\end{table}

\subsection{\label{subsec:QRPA} QRPA calculations}
We use the QRPA code described and tested in Ref.~\cite{Ter10}. The calculations are performed for $K=0$$-$$8$ and $\pi=\pm$ for convergence of the NME. The dimension of 
the two-canonical-quasiparticle space [see  Eq.~(\ref{eq:QRPA_creation_operators})] for expressing the QRPA Hamiltonian matrix is $\simeq$58,000 for $K=0, 1$ and $\simeq$10,000$-$25,000 for other $K$ values. This dimension is controlled by a pair of cutoff occupation probabilities $v_\textrm{cut}^\textrm{ph}$ and $v_\textrm{cut}^\textrm{pp}$ applied to the occupation probabilities of the canonical basis states; see Ref.~\cite{Ter10}. The dimensions of the four modes of $K=0,1$ and $\pi=\pm$ are much larger than those of the other $(K\pi)$ values because the former $(K\pi)$ modes have spurious solutions \cite{Rin80} in QRPA calculations based on the deformed mean and pair fields; the translational invariance is also broken by the nuclear wave functions. Those dimensions are determined from our experiences with separation of the spurious solutions  and convergence of sum rules in the mass region of $A\approx$150 \cite{Ter10,Ter11}. 
Smaller dimensions can be used as $K$ increases, as long as $K$ does not have spurious solutions.

The $\gamma$ vibrational solution for $^{150}$Nd appears at an energy of 1.766 MeV, whereas the experimental value is 1.062 MeV \cite{nndc,Kru02}. The calculated $B(E2;0^+$$\rightarrow$$2^+)$ is 0.0380 $e^2$b$^2$, and the experimental value is 0.069(3) $e^2$b$^2$ \cite{nndc}. The QRPA with the setup in this paper is better near the center of the well-deformed rare-earth region ($A$ $\simeq$ 164) \cite{Ter11}. However, the fact that the QRPA energy is higher than the experimental data implies that the QRPA solutions are far from the breaking point of the QRPA. To our knowledge, there are no data for the two-particle transfer strength for  
$^{150}$Nd $\rightarrow$ $^{148}$Nd or $^{150}$Sm $\rightarrow$ $^{148}$Nd, which are relevant to  Eq.~(\ref{eq:NME_closure_2n_removal_2p_addition_QRPA}).

The proton-neutron pairing energy density functional is not used in our QRPA calculations. Many QRPA approaches, e.g., \cite{Mus13}, introduce this energy density functional so as to reproduce the NME of two-neutrino double-beta $(2\nu\beta\beta)$ decay, and the NMEs of both $0\nu$ and $2\nu$ $\beta\beta$ decay are significantly reduced. The pairing correlations are generally significant only near chemical potentials; therefore, most studies of the proton-neutron pairing correlations independent of the $0\nu\beta\beta$ decays have concentrated on the $N=Z$ line and its narrow vicinity, e.g., \cite{Goo79}. The calculations of Ref.~\cite{Civ97} show that the proton-neutron pairing gap vanishes at $N-Z=6$ for $A=60$$-$70. Although we do not know of any studies of the dynamical effects of the proton-neutron pairing correlations around $A=150$ independent of double-beta decay, strong proton-neutron pairing correlations such as those of QRPA solutions reproducing the $2\nu\beta\beta$ NMEs do not seem understood easily in those nuclei with $N-Z=30$.

The spurious QRPA solutions of $K=0, 1$ ($\pi=\pm$) are not included in the sets of intermediate states. The spurious solutions emerge because the HFB ground states break the symmetries of the Hamiltonian, and the symmetries are retrieved in the QRPA order. There are no spurious states 
independent of the ground states. 

\subsection{\label{subsec:calculation_0vbb_transition_operator_me} Calculation of matrix elements of \textbf{0}$\bm{\nu\beta\beta}$ transition operator}
\begin{figure}
\includegraphics[width=6cm]{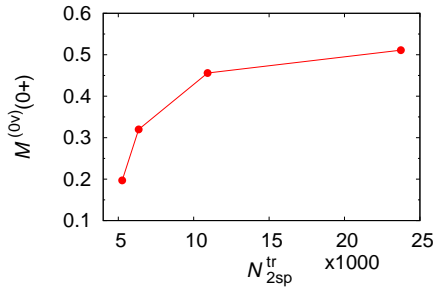}
\caption{ \label{fig:NME_2canonical_sp} (Color online) $(K\pi)=(0+)$ component of NME as a function of $N^\textrm{tr}_\textrm{2sp}$.}
\end{figure}
The two-proton and two-neutron spaces are also truncated in the calculations of the matrix elements of the $0\nu\beta\beta$ transition operator by introducing another cutoff occupation probability $v^\textrm{tr}_\textrm{cut}$ for the canonical single-particle states. 
If ($v_\mu^2$ $>$ ${v^\textrm{tr}_\textrm{cut}}^2$ or 
$v_{\mu^\prime}^2$ $>$ ${v^\textrm{tr}_\textrm{cut}}^2$) and 
($1 - v_\mu^2$ $>$ ${v^\textrm{tr}_\textrm{cut}}^2$ or 
$1 - v_{\mu^\prime}^2$ $>$ ${v^\textrm{tr}_\textrm{cut}}^2$) (the same condition is applied to another pair of $\nu \nu^\prime$), then those $\mu\mu^\prime$ and $\nu \nu^\prime$ states are used for 
$V^{(0\nu)}_{\mu\mu^\prime,\nu \nu^\prime}(\bar{E}_a)$. 
The underlying idea is that if both canonical single-particle states $\mu$ and $\mu^\prime$ are almost unoccupied or occupied, they are not used (the same idea is applied to $\nu\nu^\prime$). 

Figure \ref{fig:NME_2canonical_sp} shows the dependence of the $(K\pi)=(0+)$ component of the NME on the number of two-canonical single-particle states, $N^\textrm{tr}_\textrm{2sp}$ (summation of the numbers of the $\mu\mu^\prime$ and $\nu\nu^\prime$ states). 
Roughly speaking, the difference in $M^{(0\nu)}(0+)$ for the rightmost two points is half that for the second and third points from the right, and the $N^\textrm{tr}_\textrm{2sp}$ value of the rightmost point is twice that of the second point from the right. Extending this relation phenomenologically, we can estimate the error by truncation of our best value, that is, the rightmost point, to be around 
4\%. We use ${v^\textrm{tr}_\textrm{cut}}^2=10^{-4}$, which corresponds to $N^\textrm{tr}_\textrm{2sp} \simeq 24000$ in Fig.~\ref{fig:NME_2canonical_sp}, without any effective operator method throughout the NME calculations in this paper because the above estimated error by truncation is small. 
The factor $\mu_a$ [Eq.~(\ref{eq:mume})] of 18.51 taken from Ref.~\cite{Doi85} 
is used in our calculations. See, e.g., Ref.~\cite{Doi85} and Fig.~3 in Ref.~\cite{Hor10} regarding justifications of the closure approximation and possible values of $\bar{E}_a$.

\subsection{\label{subsec:overlap_calculations} Overlap calculations}
A truncation of the canonical quasiparticle states is introduced in Eqs.~(\ref{eq:1st-1_unnormalized_overlap}) and (\ref{eq:1st-2_unnormalized_overlap}), as in Ref.~\cite{Ter13}. The set of pairs of canonical quasiparticle states $\mu\nu$ connected by 
$j^z_\mu + j^z_\nu = K$ and $\pi_\mu \pi_\nu = \pi$ is truncated by the condition 
$v_\mu^2 > {v_\textrm{cut}^\textrm{ov}}^2$ and 
$v_\nu^2 > {v_\textrm{cut}^\textrm{ov}}^2$; those $\mu\nu$ satisfying this condition are used. 
We use ${v^\textrm{ov}_\textrm{cut}}^2 = 10^{-3}$ according to the test associated with Fig.~3 in Ref.~\cite{Ter13}.
\begin{figure}
\includegraphics[width=6.0cm]{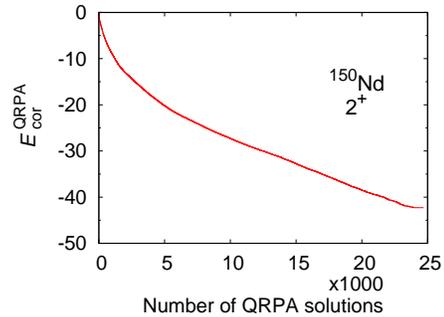}
\caption{ \label{fig:ecrqrpa_nd150_2+} (Color online) $E^\textrm{QRPA}_\textrm{cor}$ as a function of the number of QRPA solutions for $(K\pi)=(2+)$, $^{150}$Nd. At each QRPA solution number, the lowest  energy solutions are used.}
\end{figure}

To obtain a realistic NME, we need to avoid the following problem. Using all the QRPA solutions gives rise to a too-small NME of $\simeq$0.07; this is much smaller than one-tenth of the values of other groups (these values will be shown in Sec.~\ref{subsec:NME_other_groups}). The reason for this problem is that the QRPA correlations are overestimated by the Skyrme and volume pairing energy density functionals. Because these energy functionals correspond to a contact interaction, the high-momentum components do not decrease. This property causes a known problem in which the QRPA correlation energy  \cite{Fur08} 
\begin{equation}
E^\textrm{QRPA}_\textrm{cor} \simeq \frac{1}{2}\sum_a ( E_a - E^\textrm{TDA}_a),
\end{equation}
diverges \cite{Mog10}. Here, $E_a$ and $E^\textrm{TDA}_a$ denote the QRPA and Tamm--Dancoff approximation \cite{Rin80} energy of solution $a$, respectively, and the number of solutions must be the same for the two calculations. We illustrate our example of $E^\textrm{QRPA}_\textrm{cor}$ in Fig.~\ref{fig:ecrqrpa_nd150_2+}. The semi-experimental correlation energy can be defined as
\begin{equation}
E^\textrm{exp}_\textrm{cor} = E_\textrm{exp} - E_\textrm{HFB},
\end{equation}
where $E_\textrm{exp}$ is the experimental mass, and $E_\textrm{HFB}$ is the HFB energy of the ground state. $E^\textrm{exp}_\textrm{cor}$ for $^{150}$Nd is $-1.696$ MeV, and that for $^{150}$Sm is $-3.661$ MeV. In the current calculation, this problem decreases the NME unphysically because the normalization factors of the QRPA ground state are too large.

To avoid this over-correlation problem, first, we calculate the backward norms of the QRPA solutions:
\begin{equation}
\mathcal{N}^\textrm{a}_\textrm{back} = \sum_{\mu\nu} |Y^a_{-\mu-\nu}|^2.
\end{equation}
We then select the QRPA solutions that have the largest backward norms, excluding those possibly having numerical errors due to spurious components so as to reproduce the semi-experimental correlation energy, and use them to calculate the QRPA ground states used in the overlaps. 
The reason for this choice is that the backward amplitudes of many high-energy solutions are redundant, as Fig.~\ref{fig:ecrqrpa_nd150_2+} shows. This modification implies that Eq.~(\ref{eq:QRPA_ground_state}) is required only for the selected QRPA solutions, and $C^{(K\pi)I}_{\mu\nu,-\mu^\prime -\nu^\prime}$ and $C^{(K\pi)F}_{\mu\nu,-\mu^\prime -\nu^\prime}$ are changed by restricting the summation in Eq.~(\ref{eq:correlation_coefficient}). Those QRPA solutions are listed in Table~\ref{tab:QRPA_solutions_pickedup} ($K\geq 0$). $(K\pi)=(0+)$ is avoided, as mentioned above, and $|K|>4$ are not included because their backward norms are very small. All of those solutions have backward norms larger than 0.01, and their contributions to the correlation energy range from $-$0.03 MeV to $-$0.25 MeV. 
The correlation energy obtained is $-$1.721 MeV for $^{150}$Nd and $-$3.688 MeV for $^{150}$Sm.
We do not reduce the number of QRPA intermediate states.

\begin{table}
\caption{\label{tab:QRPA_solutions_pickedup} QRPA solutions selected for obtaining the QRPA ground states used in the overlaps and their properties. Note that the solutions with negative $K$ values make the same contributions to the correlation energy and the QRPA ground states.}
\begin{ruledtabular}
\begin{tabular}{ccccc}
Nucleus & $K\pi$ & $E_a$ (MeV) & $\begin{array}{cc}(E_a - E^\textrm{TDA}_a)/2 \\(\textrm{MeV}) \end{array}$& $\mathcal{N}^a_\textrm{back}$ \\
\colrule
$^{150}$Nd & 1+ & 2.263 & $-$0.061 & 0.013 \\
           & 2+ & 1.766 & $-$0.171 & 0.062 \\
           &    & 2.730 & $-$0.085 & 0.024 \\
           &    & 3.259 & $-$0.094 & 0.026 \\
           & 2$-$ & 1.984 & $-$0.094 & 0.060 \\
           &    & 2.557 & $-$0.129 & 0.050 \\
           & 3+ & 3.270 & $-$0.075 & 0.015 \\
           & 3$-$ & 3.790 & $-$0.042 & 0.015 \\
           &    & 3.897 & $-$0.046 & 0.028 \\
           & 4+ & 6.301 & $-$0.063 & 0.013 \\
$^{150}$Sm & 0$-$ & 2.829 & $-$0.149 & 0.023 \\
           &    & 3.111 & $-$0.105 & 0.017 \\
           & 1+ & 3.152 & $-$0.083 & 0.022 \\
           & 1$-$ & 2.418 & $-$0.219 & 0.014 \\
           &    & 2.858 & $-$0.136 & 0.016 \\
           & 2+ & 1.551 & $-$0.252 & 0.119 \\
           &    & 2.370 & $-$0.080 & 0.023 \\
           &    & 2.756 & $-$0.030 & 0.010 \\
           &    & 2.934 & $-$0.116 & 0.019 \\
           &    & 3.507 & $-$0.073 & 0.018 \\
           &    & 3.712 & $-$0.061 & 0.013 \\
           & 2$-$ & 1.890 & $-$0.176 & 0.132 \\
           &    & 2.461 & $-$0.091 & 0.033 \\
           &    & 2.675 & $-$0.107 & 0.031 \\
           &    & 3.399 & $-$0.075 & 0.015 \\
           & 3+ & 3.602 & $-$0.116 & 0.024 \\
           & 3$-$ & 3.202 & $-$0.054 & 0.018 \\
           &    & 3.534 & $-$0.049 & 0.056  
\end{tabular}
\end{ruledtabular}
\end{table}

\section{\label{sec:NME_output} NME}
\subsection{\label{subsec:discussion} Discussion of our NME}
\begin{figure}
\includegraphics[width=6cm]{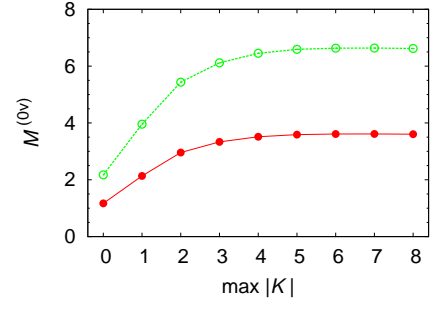}
\caption{ \label{fig:sum_nme2} (Color online) NME obtained by our method (filled circles) and 
that obtained using the HFB ground states in the calculations of the overlaps (open circles).}
\end{figure}
Figure \ref{fig:sum_nme2} illustrates the result of our NME calculation; a partial NME defined as
\begin{equation}
M^{(0\nu)} = \sum_{K^\prime=-\textrm{max}|K|}^{\textrm{max}|K|} \sum_\pi M^{(0\nu)}(K^\prime\pi)
\end{equation}
is shown as a function of max$|K|$. $(K^\prime \pi)$ indicates that of the intermediate states. The lower ones are the result obtained using our method, and the higher ones are the result of a reference calculation obtained using the HFB ground states instead of the QRPA ground states in the calculations of the overlaps. It is seen that max$|K|$ = 8 is sufficient, and our best value is 3.604. The reference value of the upper curve is 6.620, which is 84\% larger than our best value. 
  The most important information in this figure is that the QRPA correlations significantly reduce the NME through the overlaps, which are calculated using the QRPA ground states. In the reference calculation, only Eq.~(\ref{eq:0th_unnormalized_overlap}) is used with $\mathcal{N}_I = \mathcal{N}_F = 1$ in the overlap calculations.

\begin{figure}
\includegraphics[width=6cm]{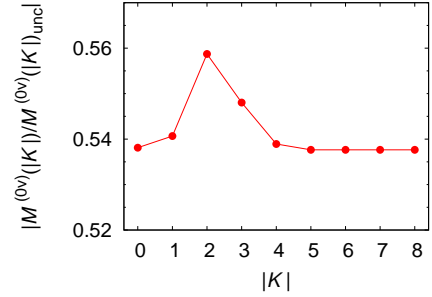}
\caption{ \label{fig:ratio_nme_nme_uncor_ov} (Color online) Ratio of $|K|$ component of the NME to that obtained using the HFB state instead of the QRPA ground state in the overlap calculations.}
\end{figure}

Figure \ref{fig:ratio_nme_nme_uncor_ov} shows the ratios of the $|K|$ components of two NMEs: $M^{(0\nu)}(|K|)$ and $M^{(0\nu)}(|K|)_\textrm{unc}$. The former is defined as
\begin{equation}
M^{(0\nu)}(|K|) = \sum_\pi M^{(0\nu)}(K\pi) \times \biggl\{ 
\begin{array}{ll}
1, K=0 \\
2, K\neq 0 
\end{array}, \label{eq:M_absK_comp}
\end{equation}
and the latter, with subscript unc indicating uncorrelated, is defined by the same equation but using the HFB ground state instead of the QRPA ground state in the overlap calculations. The relative bump indicates the effect of the QRPA correlations through the unnormalized overlap. This is seen from an approximate equation for the overlap,
\begin{eqnarray}
\lefteqn{
\frac{1}{\mathcal{N}_F\mathcal{N}_I} \langle f | \prod_{K_1 \pi_1} \exp[\hat{v}^{(K_1\pi_1)\dagger}_F] O^{F}_{a} O^{I\dagger}_a \exp[\hat{v}^{(K_1\pi_1)}_I] | i \rangle } \nonumber \\
&&\simeq \frac{1}{\mathcal{N}_F\mathcal{N}_I} \langle f | \exp[\hat{v}^{(K\pi)\dagger}_F] O^{F}_{a} O^{I\dagger}_a \exp[\hat{v}^{(K\pi)}_I] | i \rangle ,
\label{eq:overlap_approx}
\end{eqnarray}
where $K\pi$ denotes those of the QRPA state $a$. For $K>4$, this equation leads to
\begin{eqnarray}
\frac{1}{\mathcal{N}_F\mathcal{N}_I} \langle f | O^{F}_{a} O^{I\dagger}_a | i \rangle ,
\label{eq:overlap_no_v}
\end{eqnarray}
and $\langle f | O^{F}_{a} O^{I\dagger}_a | i \rangle$ is the uncorrelated overlap. 
Therefore, the flat portion of Fig.~\ref{fig:ratio_nme_nme_uncor_ov} for $|K|>4$ is equal to 
$1/(\mathcal{N}_F \mathcal{N}_I)$, and the relative bump indicates the effect of 
$\hat{v}^{(K\pi)\dagger}_F$ and $\hat{v}^{(K\pi)}_I$ in the unnormalized overlaps. This effect is 
at most a few percent at the limited $K$ values; thus, the greatest effect of the QRPA correlations through the overlaps appears in the normalization factors. 

$\mathcal{N}_F \mathcal{N}_I = 1.860$ in the calculation yielding $\mathcal{M}^{(0\nu)}=3.604$ is much larger than 1 because the contributions of many QRPA solutions are accumulated. 
According to the equations of the correlation coefficients [Eq.~(\ref{eq:correlation_coefficient})], a simplified estimation
\begin{equation}
\mathcal{N} \sim \exp\Bigl[\frac{1}{2}\sum_a \mathcal{N}^a_\textrm{back}\Bigr]  
\label{eq:normalization_estimated}
\end{equation}
is possible.
This $\mathcal{N}$ represents either $\mathcal{N}_I$ or $\mathcal{N}_F$. Using this equation and the values of $\mathcal{N}^a_\textrm{back}$ in Table~\ref{tab:QRPA_solutions_pickedup}, we obtain 
$\mathcal{N}_F \mathcal{N}_I \sim 2.4$. Therefore, the large normalization factors are not surprising.

It is worth noting again that the high-energy components of the excitations included in 
$O^{F}_a O^{I\dagger}_a$ do not contribute to the unnormalized overlap because the initial and final ground states have different proton and neutron configurations at the Fermi surfaces. Therefore, the effect of the large normalization factors manifests itself in the NME. 

We also performed a calculation using 27 ($^{150}$Nd) and 79 ($^{150}$Sm) QRPA solutions ($K\geq 0$) with the largest $\mathcal{N}_\textrm{back}^a$  so as to have twice the correlation energy for comparing the effects of the QRPA correlations in the ground states used in the overlaps. The NME of this calculation was 2.990, which is 17\% lower than our best value. Thus, low-energy solutions with large backward norms are more important than the same number of  solutions with higher energy and smaller backward norms. However, the NME does not converge as the number of QRPA solutions used for the ground states increases greatly, as mentioned in Sec.~\ref{subsec:overlap_calculations}.

\subsection{\label{subsec:NME_other_groups} NMEs of other groups}
\begin{table}
\caption{\label{tab:NME_other_groups} Our NME and those of other groups for $^{150}$Nd $\rightarrow$ $^{150}$Sm. PnQRPA denotes proton-neutron QRPA, and likeQRPA represents like-particle QRPA. IBM-2 indicates interacting boson model-2, and GCM is the generator coordinate method. The values obtained with $g_A=1.25$ or close to it are listed, if that value is available. $M^{(0\nu)}$ = 3.14 and 2.71 of
PnQRPA (Skyrme, volume pairing) were obtained with SkM$^\ast$ and modified SkM$^\ast$, respectively.}
\begin{ruledtabular}
\begin{tabular}{lcc}
Method & $M^{(0v)}$ & Ref. \\
\colrule
PnQRPA (CD-Bonn, G matrix) & 3.34 & \cite{Fan11} \\
PnQRPA (Skyrme, volume & 3.14, 2.71 & \cite{Mus13} \\
pairing)               &            &              \\
IBM-2 & 2.321 & \cite{Iac11, Bar09}\\
Projected HFB & 3.24{$\pm$}0.44 & \cite{Rat10} \\
Energy density functional & 2.190 & \cite{Vaq13} \\
(Gogny, GCM, projection) & & \\
Relativistic (GCM, projection) & 5.60 & \cite{Son14} \\
LikeQRPA & 3.604 & Current paper
\end{tabular}
\end{ruledtabular}
\end{table}
Our $M^{(0\nu)}$ value and those obtained by other groups are summarized in Table~\ref{tab:NME_other_groups}.
Only those obtained using the most current methods are listed. The major difference is the approximation for the nuclear wave functions, as noted in the table; however, there are also many other differences. 
The value of Ref.~\cite{Fan11} in the table was obtained  by a proton-neutron QRPA calculation with $g_A=1.25$ ($M^{(0\nu)}=2.55$ when   $g_A=0.94$). The interaction used was the nuclear Brueckner G matrix derived from the charge-dependent Bonn one-boson exchange potential. The overlaps were calculated using a simplified method \cite{Sim04}. The number of proton-neutron quasiparticle pairs, which determines the dimension of the proton-neutron QRPA equation, was 921 for $(K\pi)=(0+)$, $^{150}$Nd and $^{150}$Sm. 
The particle-hole matrix elements in the QRPA equation were multiplied by the factor $g_{ph}=0.90$. This factor is fitted to the experimental position of the Gamow--Teller giant resonance in the intermediate nucleus, of which the parent nucleus is $^{76}$Ge. The particle-particle matrix elements were multiplied by another factor, $g_{pp}$. This factor was fitted to the experimental value of the $2\nu\beta\beta$ decay NME of 0.07 MeV$^{-1}$ for $^{76}$Ge \cite{Bar10}. The $\bar{E}_a$ used was 7 MeV. Self-consistent Bonn-CD short range correlations were included. The finite nucleon size effects and higher-order weak currents were also included according to Refs.~\cite{Rod07,Sim08,Sim09}.

The value cited from Ref.~\cite{Mus13} was obtained by another proton-neutron QRPA calculation with $R=1.2A^{1/3}$ fm and $g_A=1.25$. The short-range correlation corrections were omitted on the basis of the suggestion from recent studies \cite{Sim09,Eng09} that the realistic short-range correlations affect the double-beta matrix elements only slightly; see also Ref.~\cite{Kor07}. The authors of this paper used the Skyrme and volume pairing energy density functionals. For the former functional, they modified the parameter set SkM$^\ast$ 
so as to reproduce the location and fraction of the observable strength of the Gamow--Teller resonance.   
The strength of the $T=0$ component of the pairing energy density functional was determined so as to reproduce an experimental $2\nu\beta\beta$ NME. The $(T=1,T_z=0)$ component was determined in such a way that the Fermi $2\nu\beta\beta$ matrix element vanished \cite{Rod11}, and the $(T=1,T_z=\pm 1)$ components were determined using the pairing gaps of the HFB calculations and the experimental gaps obtained from the mass differences. The dimension of the two-quasiparticle states in the proton-neutron QRPA equation was around 15,000. The overlaps were calculated using the HFB ground states.

The value of Refs.~\cite{Iac11,Bar09} cited in the table was obtained by a calculation of the interacting  boson model-2 (IBM-2) with $R=1.2A^{1/3}$ fm. The wave functions were provided by the proton-neutron IBM-2 \cite{Iac87}, and the formulation in Ref.~\cite{Sim99} was used for the calculation of the NME. The finite nucleon size was taken into account by replacing $g_A$ and $g_V$ with dipole forms, and the short-range correlations were included by multiplying the neutrino potential by the Jastrow function squared. 

The value of Ref.~\cite{Rat10} cited in Table~\ref{tab:NME_other_groups} is an average of eight values obtained by projected HFB calculations. The differences are the parametrization of the Jastrow-type function used for the short-range correlations and the variations in the multipole-multipole-type interactions. 
A value of $g_A=1.254$ was used ($M^{(0\nu)}$ = 3.59 $\pm$ 0.50 when $g_A$ = 1.0). The average energy of the intermediate states relative to the mean value  of the initial and final state energies was 
$1.12A^{1/2}$ MeV. The finite size effect of nucleons was introduced by a dipole form factor.

The value of Ref.~\cite{Vaq13} in Table~\ref{tab:NME_other_groups} was obtained by an energy density functional (Gogny \cite{Ber84}) method including deformation and pairing fluctuations explicitly on the same footing using the generator coordinate  method (GCM) \cite{Rin80,Ben03} with projected HFB wave functions. The neutrino potential used ($g_A=0.925$) includes the nucleon finite size effect, higher-order currents, and short-range correlations \cite{Men09,Rod10}, and the result was obtained as the sum of the Fermi and Gamow--Teller terms. 
The details of the NME calculation were based on Ref.~\cite{Men09}, in which $R=1.2A^{1/3}$ fm was used; see also Ref.~\cite{Rodr11}.

The value of Ref.~\cite{Son14} was obtained using the GCM plus projections in a relativistic framework with $g_A=1.254$ without the contributions of the short-range correlations. All of those calculations used the closure approximation. Compared to the compilation in 2013 \cite{Mus13}, the value of Ref.~\cite{Vaq13} is an updated one, and the value of Ref.~\cite{Son14} and our value are new. For other differences in the details, see the references in Table~\ref{tab:NME_other_groups}. 
The remarkable point of our result, which we can state considering the many differences in the methods  described above, is that a value close to the pnQRPA values with the similar setup is obtained without any effective methods known to reduce $M^{(0\nu)}$, e.g., the proton-neutron pairing energy density functional.

\section{\label{sec:consistency_QRPA_approach} Consistency of QRPA approach}
We have one more fundamental point to make regarding the QRPA approach: how the equality of Eqs.~(\ref{eq:NME_closure_double_beta_QRPA}) and (\ref{eq:NME_closure_2n_removal_2p_addition_QRPA}) [or (\ref{eq:NME_closure_2p_addition_2n_removal_QRPA})] can be achieved. 
Suppose that the strength of the Coulomb residual interaction was changed arbitrarily in the QRPA calculation. This change would not affect the proton-neutron QRPA solutions, whereas the like-particle QRPA solutions would be affected. Therefore, it is impossible to obtain the equality of Eqs.~(\ref{eq:NME_closure_double_beta_QRPA}) and (\ref{eq:NME_closure_2n_removal_2p_addition_QRPA}) without modifying the usual QRPA approach. Apparently the same ground-state wave functions should be used for these equations, and the only candidate is the extended QRPA ground states
\begin{eqnarray}
&& | I \rangle = \frac{1}{ \mathcal{N}_{\textrm{pn},I} \mathcal{N}_{\textrm{like},I} } \prod_{K\pi} 
\exp\bigl[ \hat{v}^{(K\pi)}_{\textrm{pn},I} \bigr] 
\exp\bigl[ \hat{v}^{(K\pi)}_{\textrm{like},I} \bigr]| i \rangle, \nonumber \\
&& | F \rangle = \frac{1}{ \mathcal{N}_{\textrm{pn},F} \mathcal{N}_{\textrm{like},F} } \prod_{K\pi} 
\exp\bigl[ \hat{v}^{(K\pi)}_{\textrm{pn},F} \bigr] 
\exp\bigl[ \hat{v}^{(K\pi)}_{\textrm{like},F} \bigr] | f \rangle. \nonumber \\ \label{eq:extended_QRPA_gs}
\end{eqnarray} 
In this subsection, we use the subscript pn to indicate the proton-neutron QRPA and the subscript like to indicate the like-particle QRPA. The method of obtaining the components of these equations does not change. Because these two types of QRPA do not have coupling (therefore, there are two QRPAs), the operators
$\exp[\hat{v}^{(K\pi)}_{\textrm{pn},I}]$ and $\exp[\hat{v}^{(K\pi)}_{\textrm{like},I}]$ commute 
with each other approximately. The same property holds for the final state. Thus, we can introduce the above product wave functions.

Let us investigate the implications of this extension. The well-established equation for calculating the transition strength between the ground and proton-neutron QRPA states is
\begin{eqnarray}
\lefteqn{ \langle a | c^\dagger_p c_n |I\rangle } \nonumber \\
&\simeq& \frac{1}{\mathcal{N}_{\textrm{like},I}^{\,2}} \prod_{K^\prime \pi^\prime} \langle i | \exp\bigl[\hat{v}^{(K^\prime \pi^\prime)\dagger}_{\textrm{like},I}\bigr] \exp\bigl[\hat{v}^{(K^\prime\pi^\prime)}_{\textrm{like},I}\bigr] | i\rangle \nonumber \\
&&\times \frac{1}{\mathcal{N}_{\textrm{pn},I}^{\,2}} \prod_{K\pi} 
\langle i | \exp\bigl[ \hat{v}^{(K\pi)\dagger}_{\textrm{pn},I}\bigr] 
O^{\textrm{pn},I}_a c^\dagger_p c_n 
\exp\bigl[ \hat{v}^{(K\pi)}_{\textrm{pn},I}\bigr] | i\rangle \nonumber \\ 
&=& \frac{1}{\mathcal{N}_{\textrm{pn},I}^{\,2}} \prod_{K\pi} 
\langle i | \exp\bigl[ \hat{v}^{(K\pi)\dagger}_{\textrm{pn},I}\bigr] O^{\textrm{pn},I}_a 
c^\dagger_p c_n 
\exp\bigl[ \hat{v}^{(K\pi)}_{\textrm{pn},I}\bigr] | i\rangle. \nonumber \\
\label{eq:pn_tr_me}
\end{eqnarray}
Thus, this calculation is not affected by the extension. The explicit QRPA ground-state wave function is known to be unnecessary for calculating this equation \cite{Rin80}.

\begin{figure}
\includegraphics[width=7.5cm]{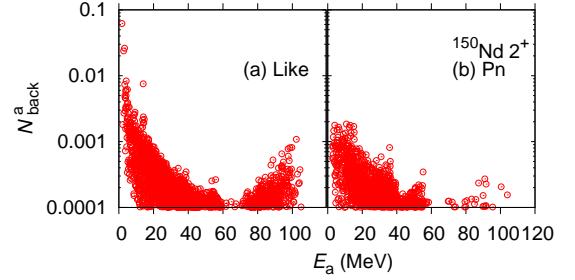}
\caption{ \label{fig:e_lnbnorm_nd150_2+_like_pn} (Color online) $\mathcal{N}^a_{\textrm{back}}$ as functions of $E_a$ of (a) the like-particle and (b) proton-neutron QRPA solutions for $^{150}$Nd, $(K\pi)=(2+)$.}
\end{figure}
As for the overlap used in Eq.~(\ref{eq:NME_closure_double_beta_QRPA}), it follows that 
\begin{eqnarray}
\langle a_F | a_I \rangle &\simeq& \frac{1}{\mathcal{N}_{\textrm{pn},F} \mathcal{N}_{\textrm{pn},I}}
\frac{1}{\mathcal{N}_{\textrm{like},F} \mathcal{N}_{\textrm{like},I}} \nonumber \\
&&\times \langle f | \exp\bigl[ \hat{v}^{(K\pi)\dagger}_{\textrm{pn},F} \bigr] 
\exp\big[ \hat{v}^{(K\pi)\dagger}_{\textrm{like},F}\bigr] O^{\textrm{pn},F}_a \nonumber \\ &&\times O^{\textrm{pn},I\dagger}_a 
\exp\bigl[ \hat{v}^{(K\pi)}_{\textrm{like},I}\bigr] \exp\bigl[ \hat{v}^{(K\pi)}_{\textrm{pn},I}\bigr] | i \rangle. \label{eq:overlap_extended}
\end{eqnarray}
The influence of the like-particle QRPA correlations cannot be removed from this equation unless the influence is negligible. The $\mathcal{N}^a_{\textrm{back}}$ values of the like-particle and proton-neutron QRPA solutions are compared in Fig.~\ref{fig:e_lnbnorm_nd150_2+_like_pn}. 
A few tens of the largest $\mathcal{N}^a_\textrm{back}$ values consist of those of the like-particle QRPA solutions. Thus, neglecting the proton-neutron QRPA correlations in the QRPA ground states [Eq.~(\ref{eq:extended_QRPA_gs})] can be justified as long as the prescription described in Sec.~\ref{subsec:overlap_calculations} is used; however, neglecting the like-particle QRPA correlations may not be a good approximation. That is, the like-particle QRPA correlations may be necessary in the overlap in  Eq.~(\ref{eq:NME_closure_double_beta_QRPA}) taking the double-beta path.

\section{\label{sec:simple_test_cal}Simplified test calculations}
In this section, we show the validity of our method by simplified calculations. One is the equivalence of the two paths of the $0\nu\beta\beta$ decay under the closure approximation shown by Eqs.~(\ref{eq:NME_closure_double_beta_QRPA}) and (\ref{eq:NME_closure_2n_removal_2p_addition_QRPA}), and another is the NME of $2\nu\beta\beta$ decay. 
\subsection{\label{subsec:eqv_2_paths}Equivalence of the two paths}
In order to demonstate this equivalence without large-scale computations we use a fictitious $0\nu\beta\beta$ decay of light nuclei $^{26}$Mg $\rightarrow$ $^{26}$Si with the box size of $\max(z) = \max(\rho) = 10$ fm and 23 B-spline mesh points per dimension. The cutoff quasiparticle energy in the HFB calcuation is 40 MeV, and the maximum $K$ quantum number of the QRPA intermediate states is 5. 

 It is possible to simplify the equation for calculation by concentrating on a demonstration of the equivalence  of the two paths. First, the factor included in the overlaps
\begin{equation}
\frac{1}{{\cal N}_{\textrm{pn},F}{\cal N}_{\textrm{pn},I}}
\frac{1}{{\cal N}_{\textrm{like},F}{\cal N}_{\textrm{like},I}}, \label{eq:the_factor}
\end{equation} 
is omitted because this factor is shared by the two equations (\ref{eq:NME_closure_double_beta_QRPA}) and (\ref{eq:NME_closure_2n_removal_2p_addition_QRPA}). Second, the effect of the QRPA correlations on the unnormalized overlaps is neglected because this effect is not significant as shown in Sec.~\ref{subsec:discussion}. Namely, the HFB ground states are used in the overlaps instead of the QRPA ground states in this subsection. We call the NME obtained under this simplification test NME for distinguishing it from the correct NME. 

We use the Skyrme energy density functional (SkM$^\ast$) and volume pairing energy density functional with the strengths 
$-150.0$ MeV fm$^3$ (protons) and $-270.0$ MeV fm$^3$ (neutrons) for $^{26}$Mg and 
$-270.0$ MeV fm$^3$ (protons) and $-200.0$ MeV fm$^3$ (neutrons) for $^{26}$Si. 
The HFB ground-state solution for $^{26}$Mg has $\Delta_p = 0.949$ MeV and $\Delta_n=3.083$ MeV, and that  for $^{26}$Si has $\Delta_p = 2.586$ MeV and $\Delta_n = 1.749$ MeV. The quadrupole deformation is $\beta = -0.118$ for both $^{26}$Mg and $^{26}$Si with a negligible difference, and the total root-mean-square radius is $\simeq$3.0 fm for both nuclei. 

 In the QRPA calculations, we use all of the two-quasiparticle states possible from the quasiparticle states obtained by the HFB calculations. The dimension of the two-quasiparticle space is, e.g. for ($K\pi$)=(0+), 8459 for the like-particle QRPA  and 8406 for the proton-neutron QRPA. Contamination of the real states by the spurious components is unavoidable in these like-particle QRPA calculations  because of the reduced computation scale. Examining the contamination by the transition strengths of the operators associated with the symmetries [e.g., the particle number operator for $(K\pi)=(0+)$], we removed a few QRPA states close to the spurious QRPA states in terms of the energy from the calculation of the test NME at each $(K\pi)$ mode having the spurious state(s). (The spurious QRPA states are always removed.)
  
  We obtained the test NME of $-5.45$ using the path of the two-neutron removal followed by the two-proton addition (the like-particle QRPA) and $-5.89$ from the double-beta path (the proton-neutron QRPA). The absolute value of the former test NME is 7.5\% smaller than that of the latter one. If the additional removal of the contaminated low-lying QRPA states is not made, the test NME of the like-particle QRPA approach is $-3.35$; the absolute value of this is 43\% smaller than that of the proton-neutron QRPA approach. The possible causes for the 7.5\% discrepancy are 
the simple removal method of the spurious components, ignoring the QRPA correlaion effect in the overlaps, and the scale of the computation: the box size and the number of mesh points. Considering these simplifications, we conclude that the equivalence of the two paths under the closure approximation is demonstrated approximately by this test calculation.

\subsection{ \label{subsec:2nubetabeta} NME of $2\bm{\nu\beta\beta}$ decay }
It is possible to obtain an approximate value of the NME of the $2\nu\beta\beta$ decay $M^{(2\nu)}$ for  $^{150}$Nd $\rightarrow$ $^{150}$Sm with the help of calculations of other group. 
The factor (\ref{eq:the_factor}) is also included in $M^{(2\nu)}$ of our method, and it is equal to $1/({\cal N}_F{\cal N}_I)=0.54$ in our calculation for $^{150}$Nd $\rightarrow$ $^{150}$Sm (see Secs.~\ref{subsec:discussion} and \ref{sec:consistency_QRPA_approach}). Therefore $M^{(2\nu)}$ of the other group calculated without the proton-neutron pairing interaction multiplied by that factor is the approximate NME of our method. 

Fortunately Ref.~\cite{Mus13} shows the $M^{(2\nu)}$ calculated using the Skyrme energy (SkM$^\ast$) and volume pairing energy density functionals for that decay instance, and the values obtained without the proton-neutron pairing interaction  are also shown in their Fig.~2(d); those values are 0.17 ($g_A=1.25$) and 0.11 ($g_A=1.0$). Multiplying them by 0.54, we obtain 0.09 and 0.06 for $g_A=1.25$ and 1.0, respectively. The latter value agrees well with the experimental value of $0.0579\pm 0.0032$; this values is obtained from the average of three experimental half-lives \cite{Bar13} (for the experiments, see Refs.~\cite{Art95,DeS97,Arg09}) and the phase space factor in Ref.~\cite{Kot12}. Here the relation between the half-life and NME in Ref.~\cite{Doi85} is used. A more recent experiment \cite{Kid14} reports $M^{(2\nu)}_\textrm{eff}=0.0465^{+0.0098}_{-0.0054}$ (see also Ref.~\cite{Bara09}). This agreement indicates the validity of our method, and it is emphasized that our QRPA solutions are not close to the breaking point of the QRPA usually encountered by the strong proton-neutron pairing interaction. This result is also consistent with our claim that the effects of the proton-neutron pairing interaction should be minor in nuclei far from the $N=Z$ line (Sec.~\ref{subsec:QRPA}).

\section{\label{sec:conclusion}conclusion and future work}
We calculated the NME for $^{150}$Nd $\rightarrow$ $^{150}$Sm using a new QRPA approach. The most remarkable point is that the overlap is calculated on the basis of the QRPA ground states that are the vacuum of quasibosons. The QRPA correlations included in the ground states were renormalized referring to the semi-experimental correlation energies. Under the closure approximation, it is possible to consider multiple virtual paths for double-beta decay. In this paper, we used the path consisting of   two-neutron removal followed by two-proton addition, and the like-particle QRPA was used to construct the intermediate nuclear states. $K$ values of up to 8  and $\pi=\pm$ were used for the intermediate states. The simplest version of the $0\nu\beta\beta$ transition operator, which contains only the Gamow--Teller and Fermi terms, was used without any effective operator methods. No finite nucleon size effect is included. We set up the calculations so as to use as large a wave function space as possible without effective methods for compensating for the truncations. The use of high-performance parallel computers is essential. This approach is inspired by the fact that the QRPA can exhaust the sum rules in practical calculations. 
The input is the Skyrme and volume pairing energy density functionals, and the proton-neutron pairing energy density functional is not used. 

A NME value was obtained which is close to the values of the other groups obtained with the similar setup. The difference is that the QRPA correlations significantly reduced the NME in our calculation because the normalization factors of the QRPA ground states were included in the overlaps. The normalization factors are included implicitly in any QRPA calculations. These factors play a special role in the NME calculation because the initial and final states are states of different nuclei. 

We argued that it was necessary to extend the QRPA ground-state wave function to a product wave function using the like-particle and proton-neutron QRPA calculations for theoretical consistency. The former QRPA solutions have larger QRPA correlations, so the current calculation is a reasonable approximation. 

 We have also checked whether the different virtual paths yield the same NME. This calculation was performed with a smaller computational scale for light nuclei due to shortage of our computational resource. We could, however, show the equivalence of the two paths approximately. The exact calculation using the proton-neutron QRPA states as the intermediate states includes the cross terms of the proton-neutron and like-particle QRPAs in the overlaps; thus, the calculation will be slightly more complicated than the calculations in this paper. 
 
 The approximate NME of $2\nu\beta\beta$ decay for $^{150}$Nd $\rightarrow$ $^{150}$Sm was obtained with the help of the result of the other group. Our method can reproduce the experimental value without the QRPA solutions close to the breaking point of the QRPA. 
 
 Calculations for decay instances that have shell model values should be performed in the future. 
It is worth noting here that the new mechanism introduced in this paper is not a special characteristic of the QRPA. 
 
\begin{acknowledgments}
 This study was supported by the HPCI strategic program Field 5, and JSPS Grants-in-Aid for Research Activity start-up under subject number 23840005 and Scientific Research(C) under subject  number 26400265. Use was made of the K computer at AICS, RIKEN through the HPCI System Research Project (hp120192 and hp120287); Mira at ALCF, ANL (0vbbqrpa); CX400 at ITC, Nagoya University through the HPCI System Research Project (z48705t); T2K-Tsukuba and Coma at CCS, University of Tsukuba through the public computer resource program for interdisciplinary research;
and SR16000 at YITP, Kyoto University.
\end{acknowledgments}
\vfill

\end{document}